\documentclass[aps,floatfix,prd,twocolumn]{revtex4}
\usepackage{graphicx}
\usepackage{dcolumn}
\usepackage{bm}
\usepackage{amsmath}
\usepackage{color}
\begin{document}

\def\dq{\frac{d^3q}{(2\pi)^3}\,}
\def\be{\begin{equation}}
\def\ee{\end{equation}}
\def\X{X(5568)}
\def\b{\bar b}
\def\q{\bar q}
\def\dd{\bar d}
\def\u{\bar u}
\def\s{\bar s}
\def\K{\bar K}
\def\Q{\bar Q}
\def\*{^{(*)}}
\def\>{\rangle}
\def\subd{su\b\dd}
\def\bit{\begin{itemize}}
\def\eit{\end{itemize}}
\def\MeV{\textrm{ MeV}}
\def\d0{D\O{}}

\title{Interpreting the $X(5568)$}

\author{T.J. Burns}
\affiliation{Department of Physics, Swansea University, Singleton Park, Swansea, SA2 8PP, UK.}
\author{E.S. Swanson}
\affiliation{
Department of Physics and Astronomy,
University of Pittsburgh,
Pittsburgh, PA 15260,
USA.}

\date{\today}

\begin{abstract}
A variety of options for interpreting the \d0 state, $X(5568)$, are examined. We find that threshold, cusp, molecular, and tetraquark models are all unfavoured. Several experimental tests for unravelling the nature of the signal are suggested.
\end{abstract}

\maketitle 

\section{Introduction}

The $\X$, recently discovered at \d0 \cite{D0:2016mwd}, is the first candidate for a hadron with four distinct quark flavours, $\subd$. Its reported mass and width,
\begin{eqnarray*}
M&=&5567.8\pm2.9^{+0.9}_{-1.9}\MeV,\\
\Gamma&=&21.9\pm6.4^{+5.0}_{-2.5}\MeV,
\end{eqnarray*}
assume the two-body decay $B_s\pi^+$ in S-wave, which implies it is a $0^+$ state. Another possibility is the decay chain $B_s^*\pi^+$, where the radiative decay $B_s^*\to B_s\gamma$ produces an undetected photon; in this case $\X$ is $1^+$ and its mass is larger than the above by the  mass difference $B_s^*-B_s=48.6^{+1.8}_{-1.6}$~MeV.

While $\X$ joins a growing number of exotic states discovered in recent years \cite{Briceno:2015rlt,Chen:2016qju}, in this paper we will argue that, even by recent standards, it is a very unusual state. Among the diverse range of explanations which have been applied to other states, none seems a natural fit for $\X$. Already several proposals have been advanced, and we comment further on these below. 

In Sec.~\ref{weak} we consider various weak coupling scenarios. The most prosaic possibility is that $\X$ is a threshold enhancement (Sec.~\ref{threshold}), arising from competition between the rapid growth in rate as phase space opens up, and rate suppression due to hadronic overlaps. The idea offers a natural explanation for peaks above two-body thresholds \cite{Swanson:2015bsa}, but in the case of $\X$ it does not fit the data.

In the cusp scenario (Sec.~\ref{cusp}) sharp features arise due to singularities in loop diagrams. It offers a viable explanation \cite{zc}, recently supported by lattice QCD~\cite{Ikeda:2016zwx}, for the $Z_b$ and $Z_c$ states, and more recently has been applied to the $P_c$ states \cite{Guo:2015umn,Liu2015}. We consider $B_s^*\pi\to B_s\pi$ rescattering, as in ref. \cite{Liu:2016xly}, and although we are able to fit the data well, this requires unnatural parameters, and in any case, we do not expect scattering in this channel to be significant.

{ We also consider (Sec.~\ref{molecular}) the molecular hypothesis.  Binding via pion exchange offers a natural explanation for states with masses slightly below relevant two-body thresholds, such as $X(3872)$ \cite{Tornqvist:2004qy,Swanson:2004pp,Thomas:2008ja}, $Y(4260)$ \cite{Close:2009ag} and $P_c(4450)$ \cite{Yang2011,Karliner2015,He2015,Burns:2015dwa}. This does not work for $\X$, which is hundreds of MeV below any such thresholds. Ref.~\cite{Xiao:2016mho} discussed the phenomenology of $\X$ as a $B\K$ molecule, but did not explain the required deep binding, which does not arise in QCD sum rules \cite{Agaev:2016urs} or in models based on vector meson exchange \cite{Zhang:2006ix}. Coupled-channel dynamics offer more possibilities. Indeed a state with the exotic flavour quantum numbers of $\X$ had been predicted using chiral Lagrangians \cite{Kolomeitsev:2003ac}, albeit with mass some 180~MeV higher than that observed. We consider a molecule arising due to the $B\K\to B_s\pi$ coupling via quark exchange, finding a potential which is attractive, but not strong enough to form the desired state. 
}


In Sec.~\ref{tetraquark} we consider the more exotic tetraquark explanation, beginning (Sec.~\ref{mass}) with the question of the mass. Tetraquark models have been widely applied to all of the exotic mesons mentioned above, and many more besides, and most recently have been applied to $\X$~\cite{Agaev:2016mjb,Chen:2016mqt,Wang:2016mee,Zanetti:2016wjn,Liu:2016ogz,Wang:2016tsi,Stancu:2016sfd}. We make several simple estimates and find that $\X$ is  too light to be a plausible tetraquark candidate. 

Setting aside the difficulty with the mass, we explore the phenomenology of the tetraquark scenario. Unlike the various weak coupling scenarios, as a tetraquark $\X$ would be accompanied by a pair of neutral partners in the same mass region (Sec.~\ref{neutral}). Depending on isospin mixing there would either be a degenerate state in $B_s\pi^0$ and very narrow partner stable to strong decay, or a pair of states in $B_s\pi^0$, one heavier, one lighter, and both narrower than $\X$.

There would also be a proliferation of other partners, both isovector and isoscalar, with different spin (Sec.~\ref{other}). We find characteristic differences in the spectra for quark and diquark models, but a feature common to both is that the lightest states are an approximately degenerate $0^+/1^+$ pair: hence regardless of whether $\X$ is itself scalar or axial, it must have a further nearby partner. Additional heavier partners are also expected, and due to a lack of strong decay channels most of these partners would be remarkably narrow.

\section{Weak Coupling Scenarios}
\label{weak}

In principle it is possible the $X$ signal arises due to a variety of weak couplings effects. For example, the structure could be a weakly bound resonance in analogy with the deuteron or the $X(3872)$. Alternatively, the signal could be due to simple ``kinematical" effects, such as a threshold enhancement or a cusp effect. We examine these scenarios in turn.

\subsection{Threshold Effects}
\label{threshold}
\begin{figure}[ht]
\includegraphics[width=8cm,angle=0]{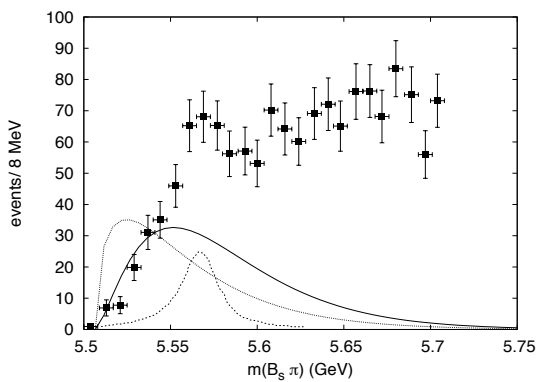}
\caption{\d0 data (points), extracted resonance signal (dashed line), and threshold models.}
\label{fig-1}
\end{figure}

Rate enhancements often appear near hadronic thresholds because (endothermic) processes behave as $(\sqrt{s}-M_C-M_D)^{1/2 + L}$ where particles $C$ and $D$ appear in the final state with relative angular momentum $L$. This sharp rise is then attenuated over a scale $\Lambda_{QCD}$ due to overlaps of the relevant hadronic wavefunctions. Such effects are ubiquitous in hadronic physics~\cite{Swanson:2015bsa}. Figure \ref{fig-1} displays the (uncut) \d0 data (points), the resonance signal extracted by \d0 (dashed line), and two model threshold effects. The dotted line is an S-wave model chosen to peak near 5568 MeV. This exhibits the characteristic fast rise, in apparent contradiction with the data. We remark that the scale used to attenuate the reaction was chosen to be substantially smaller than usual~\cite{Barnes:2003dg} in an attempt to fit the observed signal. The solid line displays the analogous P-wave model, which also does not fit the data well. Because a similar pattern holds for all waves, we conclude that it is unlikely that the $X$ signal can be explained as a threshold enhancement.

\subsection{Cusp Effects}
\label{cusp}

It is well-known that loop (such as bubble or triangle) diagrams have singularities that can generate sharp features in relevant processes~\cite{bub}. Generically, this is important when the production mechanism does not couple directly to the final state; rather the coupling is via higher mass intermediate states. This scenario provides a likely explanation of the $Z_b$ and $Z_c$ states~\cite{zc}.

In the case of the $X(5568)$ the only nearby two-particle state is $B_s^*\pi$ at 5555 MeV. We therefore postulate a generic production process that gives rise to $B_s^*\pi^\pm$ and rescatters into $B_s\pi$. The dynamics is approximated via nonrelativistic contact interactions with a Gaussian form factor dominated by a scale of order $\Lambda_{QCD}$.  

\begin{figure}[ht]
\includegraphics[width=8cm,angle=0]{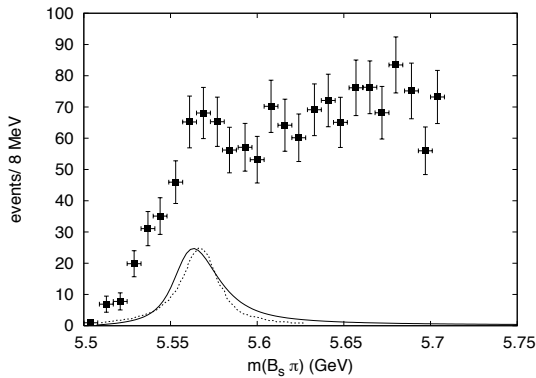}
\caption{\d0 data (points), extracted resonance signal (dashed line), and cusp model (solid line).}
\label{fig-2}
\end{figure}

Specifically

\be
\sigma \propto s\, E_\pi\,E_{B_s}\frac{p_f}{p_i} \left|\frac{\Pi(s)}{1 - \lambda \Pi(s)}\right|^2
\label{eq1}
\ee
with
\be
\Pi(s) = \int \frac{d^3q}{(2\pi)^3} q^{2\ell} \frac{\rm{e}^{-2q^2/\beta^2}}{\sqrt{s} - m_{B_s^*} - m_\pi - q^2/(2 \mu) + i \epsilon}.
\ee
Here $\mu$ is the reduced $B_s^*\,\pi$ mass and $\ell=1$. The scale $\beta$ was adjusted to fit the $X$ signal. The denominator in Eq. (\ref{eq1}) accounts for $B_s^*\pi \to B_s^*\pi$ rescattering; we see no evidence for this and set $\lambda=0$.  The result is shown as a solid line in Fig. \ref{fig-2}, where a good fit to the \d0 signal is evident. We remark that the generic features displayed here will also hold in the case of a production mechanism that proceeds via a triangle diagram.

In spite of this success, we do not regard it as likely that rescattering via $B_s^*\pi$ is a viable explanation of the \d0 signal. Firstly, this mechanism requires rescattering in P-wave, which is typically too weak to generate large effects. Furthermore, the scale required to reproduce the Breit-Wigner of width 22 MeV is $\beta = 50$ MeV. This is an order of magnitude smaller than typical scales in these applications~\cite{zc}. Finally, the process $B_s^* \pi \to B_s\pi$ is unusual because it does not entail flavour exchange, which typically must occur in low energy hadronic scattering~\cite{qex,qex2}.  In fact, it is more natural to couple the $B_s\pi$ system to $B\K$, which would generate a $J^P=0^+$ cusp slightly above 5770 MeV.

If the $B_s^*\pi$ cusp mechanism were valid it predicts a ``state" slightly above 5555 MeV (we obtain 5562 MeV) with the quantum numbers $J^P = 1^-$. Furthermore, a neutral $B_s\pi^0$ state should exist at 5557 MeV (or rather 5 MeV below the $X$) with the same width and shape as the $X$. Finally, one might also expect an analogue $B_sK$ state slightly above $B_s^*K$ (5909 MeV).

\subsection{The Molecular Hypothesis}
\label{molecular}
Candidates for molecular states typically involve meson pairs in S-wave with mass somewhat above the observed signal.  In this case no viable pairs are available and one is forced to speculate on a wider scale than normal.  For example, it is possible for the $B_s\pi$ system to scatter into $B\K$ via either quark exchange or $K^*$ exchange. If the effective potential that describes this interaction has an attractive region near the origin with a repulsive region at somewhat larger distances then it is possible that a resonance of the Gamow-Gurney-Condon (GGC) type is generated above $B_s\pi$ threshold. This scenario thus relies on some unusual S-wave dynamics and on the shape of the repulsive peak being appropriate to generating a width of 20 MeV due to tunneling.

\begin{figure}[ht]
\includegraphics[width=8cm,angle=0]{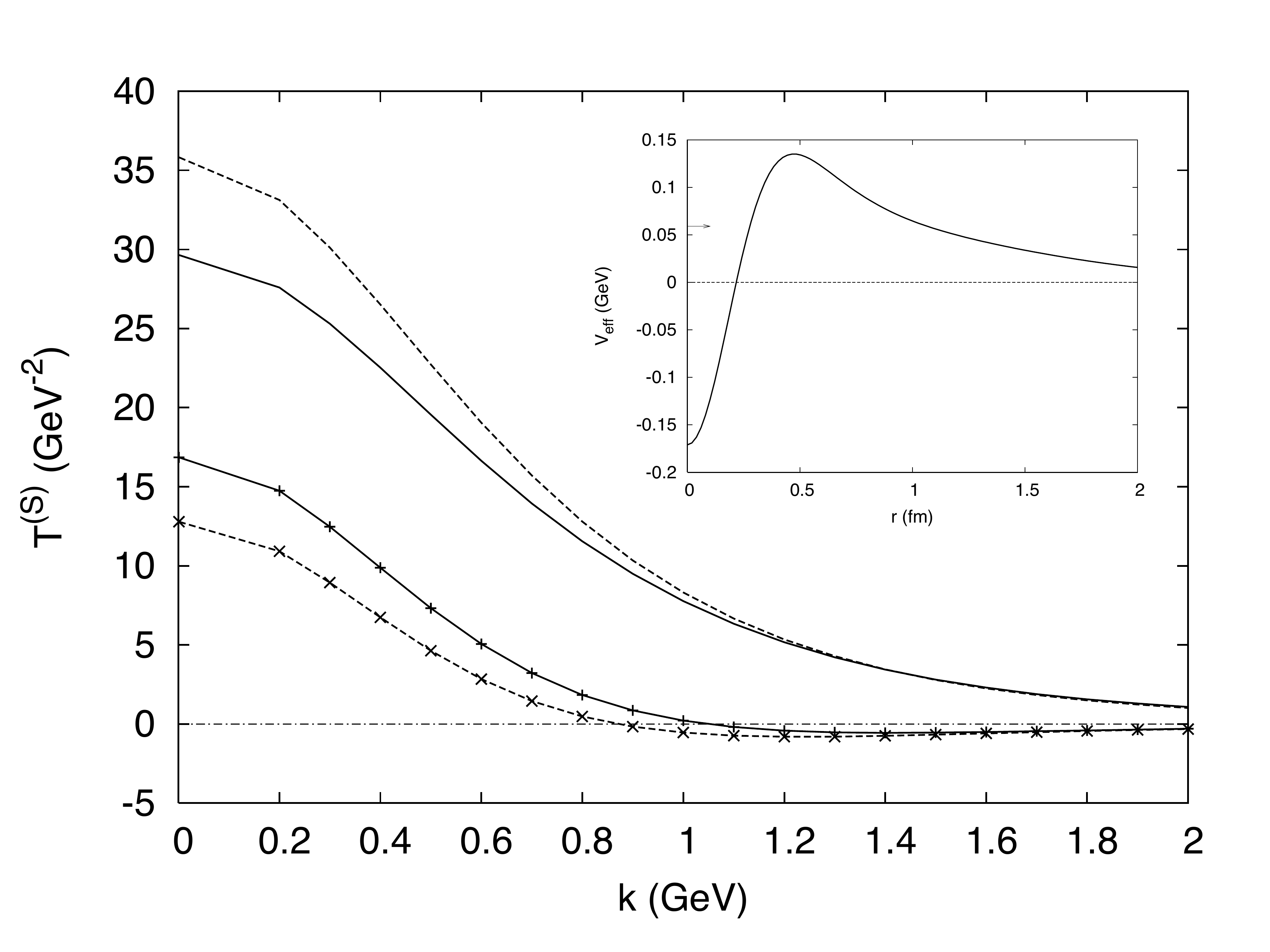}
\caption{Quark Model S-wave amplitudes for $B_s \pi \to BK$ scattering. Hyperfine prior (solid line), Hyperfine post (dashed line), confinement prior (solid line with points), confinement post (dashed line with points).
Insert: the extracted effective potential.}
\label{fig-3}
\end{figure}

We have tested the feasibility of this mechanism by computing the amplitude for $B_s\pi \to B\K$ scattering (due to quark exchange) in the nonrelativistic quark model. Our calculation employed the formalism given in ref.~\cite{qex}, with results shown in Fig.~\ref{fig-3}. The main figure shows the resulting S-wave scattering amplitude contributions from the confining (``Cornell") and hyperfine interactions. These are computed in the ``prior" and ``post" formalisms, where the interaction is defined with respect to the initial or final scattering states respectively. The two approaches should agree in the limit of accurate wavefunctions, hence the good agreement shown indicates a reasonably robust computation. 

The insert shows the equivalent S-wave potential. Surprisingly, this is precisely of the form required to produce a GGC resonance (the location of the $X(5568)$ is shown with an arrow). Unfortunately, this potential is not strong enough to generate the desired resonance. Increasing the strength of the potential eventually yields a bound state below 5507 MeV, which is, of course, not the desired result. 

Because of these observations, we do not regard the GGC resonance idea as a likely explanation of the $X$ signal -- too many delicate features would have to be realised for it to be viable.

\section{Tetraquark scenario}
\label{tetraquark}
Due to the difficulty in explaining $\X$ as a kinematic effect, we now consider the more exotic tetraquark interpretation. We find the $\X$ mass unexpectedly light for a tetraquark candidate, and show that it should be accompanied by several very narrow partners with different isospin and spin.

\subsection{Mass}
\label{mass}

Following the $\X$ discovery (we are not aware of any predictions) there have been several calculations of the mass of an $su\b\dd$ tetraquark.  Mass estimates from QCD sum rules are remarkably consistent with experiment~ \cite{Agaev:2016mjb,Chen:2016mqt,Wang:2016mee,Zanetti:2016wjn}, while those of quark models \cite{Liu:2016ogz,Wang:2016tsi,Stancu:2016sfd} are in the right region. The success of these approaches is surprising because, as we now show, according to simple arguments the $\X$ appears remarkably light for an $su\b\dd$ tetraquark.

Firstly, note that the  $bsu$ baryons $\Xi_b$ and $\Xi_b^*$ have masses of 5794 and 5945 MeV. It would be remarkable if an $su\b\dd$ tetraquark, which contains an additional valence quark, were hundreds of MeV lighter. 

Another surprise is the proximity of $\X$ to the $B_s\*\pi$ thresholds. This does not seem natural given that the tetraquark does not benefit from the chiral symmetry which protects the lightness of the pion. Instead we would expect the natural mass scale for an $\subd$ tetraquark to be near thresholds for other meson pairs with the same quark content, such as $B\*\K\*$ (whose spin-averaged mass is above 6 GeV). 

To quantify this statement, we consider the Hamiltonian of refs. \cite{Liu:2016ogz, Wang:2016tsi,Stancu:2016sfd},
\begin{equation}
H=\sum_k m_k+\sum_{ij}\alpha_{ij}\mathbf{S}_i\cdot\mathbf{S}_j,
\label{chromo}
\end{equation}
where $m_k$ is the mass of a constituent quark~\cite{Liu:2016ogz,Stancu:2016sfd} or diquark~\cite{Wang:2016tsi}, and the coefficient $\alpha_{ij}$, which depends on the color configuration of the fermion pair $ij$, is extracted from experiment and scales inversely with quark masses.  (In models such as ref.~\cite{Stancu:2016sfd}, $\alpha_{ij}$ is an operator which mixes states with different internal color configurations.) This Hamiltonian reproduces the masses and spin splittings of ordinary mesons and baryons  remarkably well~\cite{De_Rujula:1975ge,Karliner:2003sy,Karliner:2011yb}. Note that ref.~\cite{Wang:2016tsi}, following most previous diquark models \cite{Maiani:2004vq,Drenska:2009cd,Ali:2009pi}, include in the second term interactions between all pairwise combinations of fermionic constituents. In this case the idea of diquarks as effective degrees of freedom no longer seems appropriate; we comment further on this below.

 Before discussing the spin-dependent term, whose contribution varies significantly for different models, we attempt a rough estimate of the tetraquark mass on the basis of the first term, working with a constituent quark (rather than diquark) model. We take our parameters from conventional hadrons, and by inverting equation (\ref{chromo}) obtain the sum of constituent masses in a meson from the spin-averaged mass $(3M_V+M_P)/4$ of the vector ($V$) and pseudoscalar ($P$) mesons. This gives  two independent estimates for the sum of the masses of the $\subd$ constituents, considering the combinations $(u\b)(s\dd)$ and $(s\b)(u\dd)$,
\begin{eqnarray}
\frac{1}{4}\left(3B^*+B +3K^*+K\right)&=&6107\textrm{ MeV},\\
\frac{1}{4}\left(3B_s^*+B_s +3\rho+\pi\right)&=&6019\textrm{ MeV}.
\end{eqnarray}
The first of these should be a better estimate of the true masses, since the lightness of the pion has more to do with chiral symmetry than the spin-dependent interactions responsible for the splittings of other mesons. For comparison, the masses of ref. \cite{Zhao:2014qva}, obtained from averaging over different combinations of mesons to those above, yield a similar result,
\begin{equation}
\sum_k m_k=6146 \textrm{ MeV}.\label{cm}
\end{equation}
These estimates should be considered as lower limits. Fits to the spectra of baryons rather than mesons yield larger constituent masses \cite{Karliner:2003sy,Karliner:2003dt,Guo:2011gu} whose sums exceed those quoted above by hundreds of MeV.

On the basis of these estimates, the $\X$ is much lighter than would be expected as a tetraquark.

Tetraquark models for other exotic states do not encounter the same problem. In particular, since $X(3872)$, $Z_c(3900)$, $Z_c(4025)$, $Z_b(10610)$ and $Z_b(10650)$ are close to  $D^*\bar D\*$ and $B^*\bar B\*$ thresholds, it is automatic that, in tetraquark models for these states, the spin-averaging procedure analogous to the above will yield total quark masses near to the physical masses. The situation for $\X$ is markedly different, suggesting that if it is indeed a tetraquark state, it cannot easily be accommodated in the same models applied to these other putative tetraquarks.

Given the above general arguments, it is surprising that the estimates of refs. \cite{Liu:2016ogz,Wang:2016tsi,Stancu:2016sfd} are comparable to the $\X$ mass. We now discuss these estimates in more detail.

Liu \textit{et al.} \cite{Liu:2016ogz} use the Hamiltonian (\ref{chromo}) with quark (rather than diquark) constituents, with coefficients $\alpha_{ij}$ extracted from meson spectra. They obtain two scalar  $su\b\dd$ tetraquarks in the appropriate mass region, one slightly heavier, and one lighter, than $\X$. The reason for their surprisingly low masses is the chosen constituent quark masses, whose sum is much less than our estimates above,
\begin{equation}
\sum_k m_k=5700\textrm{ MeV.}
\end{equation}
The masses are taken from their earlier paper \cite{Liu:2004kd}, in which the $u$, $d$, $s$ and $c$ masses appear to have been chosen to reproduce the mass of $D_{sJ}(2632)$ in a tetraquark model, and the $b$ mass is in turn estimated from the $c$ mass. As a check on these values we use them to estimate the masses of some conventional hadrons with similar quark content, and find that they lead to drastic under-estimates, for example predicting 5250~MeV for the centre of mass of $bdu$ baryons  (compared to the experimental values $\Lambda_b=5620$~MeV, $\Sigma_b=5811$~MeV, $\Sigma_b^*=5832$~MeV), 5390~MeV for $bsu$ baryons ($\Xi_b=5794$~MeV, $\Xi_b^*=5945$~MeV),  4940~MeV for $b\bar d$ mesons ($B=5280$~MeV, $B^*=5325$~MeV), 5080~MeV for $b\bar s$ mesons ($B_s=5367$~MeV, $B_s^*=5415$~MeV). Note that tetraquark models for other states do not encounter this problem; applying the same Hamiltonian to the $Z_c$ states~ \cite{Zhao:2014qva}, one of the authors of ref.~\cite{Liu:2016ogz} used the quark masses who sum is quoted in (\ref{cm}) above.

Wang and Zhu \cite{Wang:2016tsi} use the Hamiltonian (\ref{chromo}) but with diquark constituents, and with coefficients $\alpha_{ij}$ taken from previous literature on tetraquarks. Their scalar $su\b\dd$ tetraquark has mass 5708 MeV, somewhat too heavy for $\X$, but not too far off. The comparatively low value is primarily due to the chosen diquark masses which, as with the quark masses of ref. \cite{Liu:2016ogz}, are chosen not with reference to conventional hadrons,  but from other tetraquark models. The  $bd$ mass of 5249 MeV is obtained by fitting $Y_b(10980)$ as a P-wave tetraquark~\cite{Zhu:2015bba} (see also \cite{Ali:2009pi}), and the $us$ mass of 590 MeV is from a tetraquark fit for $a_0(980)$ \cite{Maiani:2004vq}. These diquark masses are considerably lighter than those obtained in other approaches based on conventional baryons. In the model of Ebert {\it et al.} the (spin-averaged) masses of $bd$ and $us$ diquarks are 5376 MeV and 1039 MeV~\cite{Ebert:2005nc,Ebert:2005xj}. (See also ref.~\cite{Ebert:2007nw} for a comparison with other approaches, which give similar values.)

Stancu \cite{Stancu:2016sfd} employs the Hamiltonian (\ref{chromo}) with quark constituents, and the mass obtained is in good agreement with experiment. Unlike in refs \cite{Liu:2016ogz,Wang:2016tsi}, the low mass is not due to the constituent masses, whose sum is not much less than the lower bounds estimated above,
\begin{equation}
\sum_k m_k=6090\textrm{ MeV,}
\end{equation}
but is instead due to large spin splitting of {$-560$~MeV}. By comparison, the lightest $\subd$ tetraquarks in the other approaches experience splittings $-131$~MeV \cite{Wang:2016tsi} and $-225$~MeV \cite{Liu:2016ogz}. Some enhancement in the splitting is to be expected, since ref.~\cite{Stancu:2016sfd} includes all color combinations (unlike ref.~\cite{Wang:2016tsi}) and allows for full mixing across the basis states (unlike ref.~\cite{Liu:2016ogz}). However the more significant effect is the choice of coefficients $\alpha_{ij}$ in the spin-dependent term. 

In particular, for the $u\bar d$ interaction (which is the dominant contribution to the $\subd$ splitting)  the coefficient  is chosen to reproduce the $\rho-\pi$ mass difference. As remarked earlier, the lightness of the pion is not solely due to the spin-dependent interactions which control the spectra of other hadrons, so this value is likely to be an overestimate. To avoid this problem, other authors choose to extract coefficients from baryons, rather than mesons, leading to smaller values.

To check the sensitivity of the results of ref. \cite{Stancu:2016sfd}, we have reproduced the calculation with different parameter sets. Replacing the coefficients for $su$, $s\bar d$ and $u\bar d$ interactions with those of ref. \cite{Hogaasen:2004pm}, the splitting reduces to $-401$~MeV, pushing the total mass up to 5689~MeV. (Remember that the chosen quark masses are already somewhat lighter than the lower bounds quoted above.) Going to the heavy quark limit (switching off any pairwise interactions with $b$) we find $-357$~MeV, consistent with the previously quoted result for the $su\bar d$ combination   \cite{Hogaasen:2004pm}. Alternatively, using the full parameter set of ref. \cite{Liu:2016ogz}, we obtain  $-355$~MeV. 

To summarise, among the various approaches that of ref. \cite{Stancu:2016sfd} seems most promising, but we find that it can only reproduce the $\X$ mass with a choice of low quark masses and large spin coupling coefficients.

\subsection{Neutral partners}
\label{neutral}

Setting aside the apparent difficulty of its mass, we now explore some implications of the tetraquark interpretation for $\X$. Foremost among these is the existence of several narrow partner states. We restrict our discussion to states with flavour $sq\b\q$ (where $q$ is $u$ or $d$). A proliferation of partner states with flavours $qq\b\q$ and $ss\b\s$ is also expected and will not be discussed here; see refs. \cite{Liu:2016ogz,He:2016yhd}.

As a tetraquark the $|I,I_3\>=|1,\pm 1\>$ state $\X$ would have two neutral partners nearby, either isospin eigenstates $|1,0\>$ and $|0,0\>$, or linear combinations thereof. This distinguishes it from the cusp scenario, which has only one neutral state. (A third neutral state is possible due to $ss\b\s$, but we do not discuss this; most of the conclusions below are not affected.) 

If isospin is a good quantum number, the $|1,0\>$ state decaying into $B_s\pi^0$ would be degenerate with $\X$, unlike in the cusp scenario in which the $B_s\pi^0$ peak would be a few MeV lower in mass than the $B_s\pi^\pm$ peak. A more drastic consequence is that the $|0,0\>$ state would be remarkably narrow, as it has no open strong decay channels: the lowest relevant isoscalar threshold is $B\K$, more than 200 MeV heavier. The isoscalar counterpart to $\X$ could only decay by isospin violation  (into $B_s\*\pi$), radiatively, or weakly. Such a narrow state would be a striking signature for tetraquarks: as there are no isoscalar thresholds nearby, kinematic effects are unlikely to be relevant.

If instead the physical states are admixtures of $|1,0\>$ and $|0,0\>$, mixing would drive their masses apart compared to the unmixed masses, so that one is heavier, and the other lighter, than the observed $|1,\pm 1\>$ state $\X$. Their strong decays proceed via their isovector components, which are suppressed by mixing angles, so the states would be narrower than $\X$ before small phase space differences.

Experimental analysis of the $B_s\pi^0$ channel would therefore be revealing. A peak at the mass of $\X$ would indicate the possibility of tetraquark degrees of freedom (since a peak due to a cusp would be lower), and since this implies a state of pure isospin, there would have to be an extremely narrow partner state in the same mass region, which may or may not also be visible in $B_s\pi^0$. Alternatively, the observation of a pair of peaks, narrower and displaced either side of the $\X$ mass, would also indicate tetraquarks, in this case of mixed isospin.

\subsection{Other partners}
\label{other}
A tetraquark $\X$ would also have other partners (both isovector and isoscalar) with various spin quantum numbers. The proliferation of partners is a generic feature of tetraquark models, and in some cases the experimental absence of partners can be understood as a result of their being so broad as to be effectively unobservable \cite{Burns:2004wy,Burns:2010qq}. We will see that this does not apply to the partners of $\X$.

Assuming S-wave constituents, for each flavour there are two scalars ($0^+$), three axials ($1^+$) and a single tensor~($2^+$)  {in the diquark-antidiquark picture}. In the most general models, the multiplicity of states doubles again, due to the two color combinations $(qq)^{\overline 3}(\bar b\bar q)^{3}$ and $(qq)^{6}(\bar b\bar q)^{\overline 6}$. Diquark models typically ignore the second combination, thus halving the total number of states, although this may not be justified \cite{Stancu:2006st,Brink:1994ic}. Of the models applied to $\X$, ref.~\cite{Wang:2016tsi} belongs to this second class of models with a truncated spectrum, whereas refs~\cite{Liu:2016ogz,Stancu:2016sfd} include all color combinations and so predict twice as many states. (As noted previously, refs~\cite{Liu:2016ogz,Stancu:2016sfd} differ in the treatment of mixing of internal color configurations, but the total number of states is the same.) For much of our discussion we refer to the truncated spectrum, although many of our conclusions are easily generalised to the full spectrum.

\begin{figure*}[ht]
\includegraphics[width=7cm,angle=0]{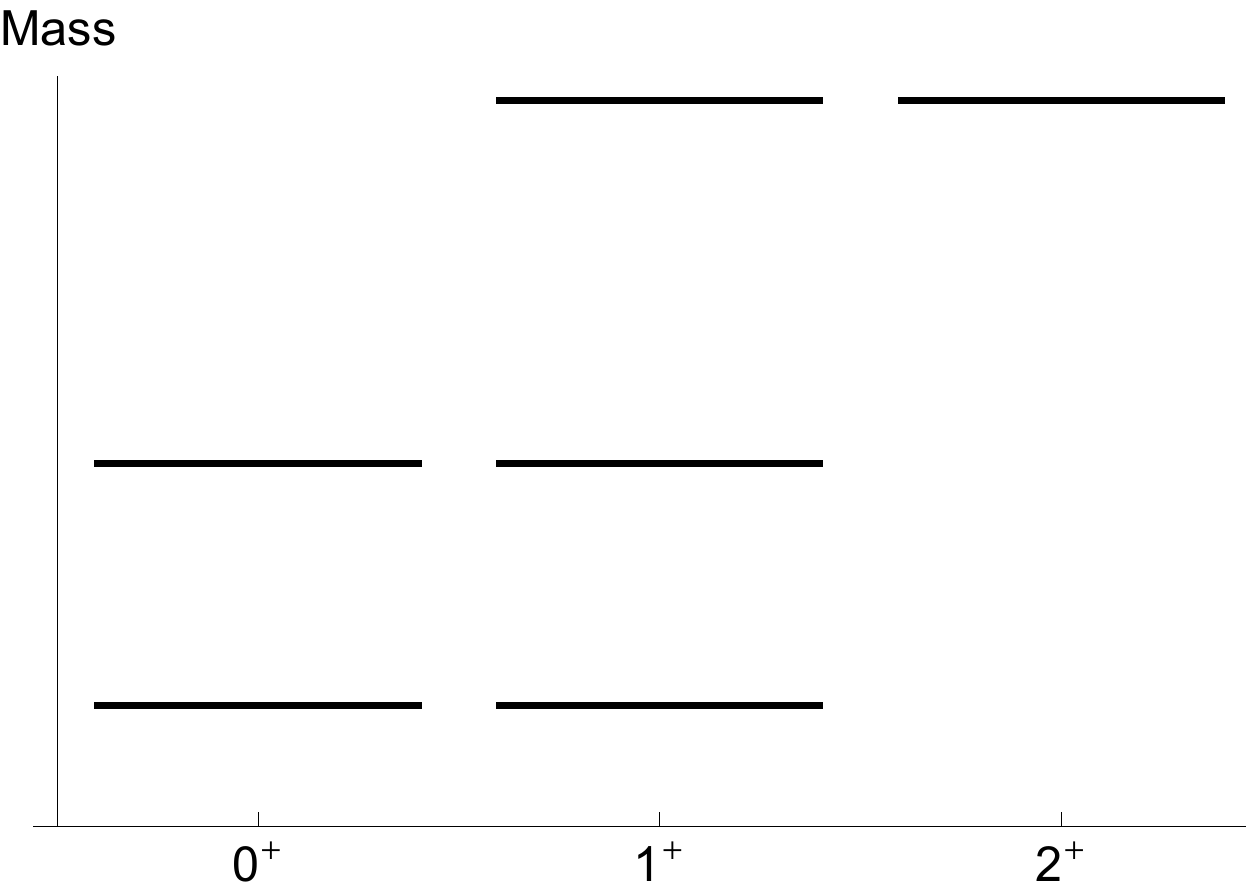}
\qquad\qquad
\includegraphics[width=7cm,angle=0]{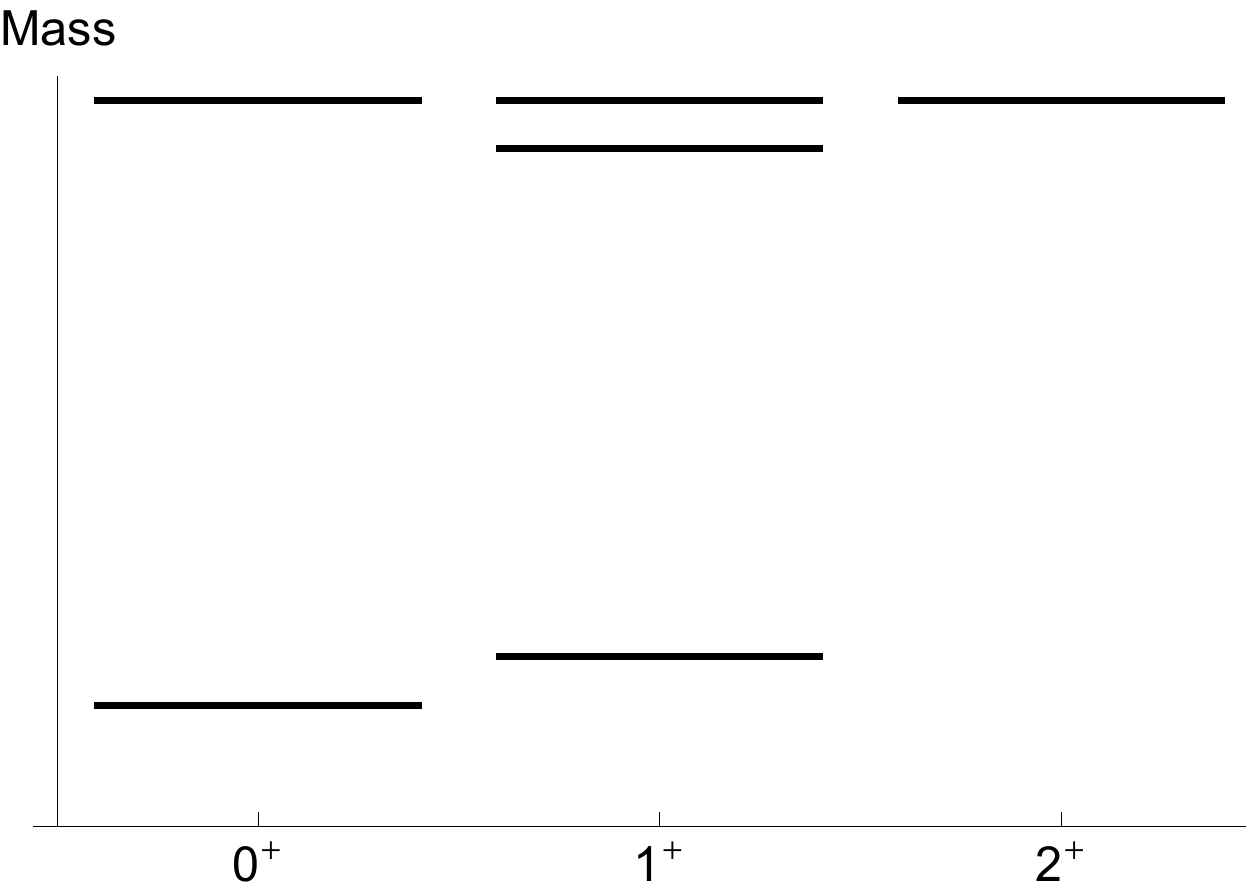}
\caption{The mass level ordering of $qq\b\q$ tetraquarks, in arbitrary units, for the truncated spectrum with half as many color combinations as the most general models. The left plot shows the spectrum in models (such as ref.~\cite{Wang:2016tsi}) with pairwise spin-spin interactions among all quark constituents{, in the heavy quark limit. The ordering of the degenerate doublets is as shown, but the spacing between doublets depends on model parameters. In this plot the spectra are given by equations~(\ref{diq:1})--(\ref{diq:3}) choosing $\kappa_{q\bar q}=\kappa_{qq}/2$, as described in the text. The right plot shows the spectrum for models (such as ref.~\cite{Ali:2016gdg}) with spin-spin interactions only within diquarks. As shown in equations (\ref{mass:1})-(\ref{mass:4}), the mass splittings are controlled by two parameters, with a small gap separating the lightest $0^+$ and $1^+$, an additional higher-lying $1^+$, and higher still a degenerate $0^+$, $1^+$, $2^+$ triplet.}}
\label{fig-spectrum}
\end{figure*}

{ The models are further distinguished according to whether in equation (\ref{chromo}) the spin-dependent interactions act pairwise on each of the quark constituents, or are restricted to acting ``within'' the diquarks. We consider these different models in turn.

For the first type of model, with pair-wise interactions among all quark constituents, note that since $\alpha_{ij}$ scales inversely with quark masses, the hyperfine terms involving the $\bar b$ quark are strongly suppressed. In the heavy quark limit (setting these  terms to zero), the mass splittings among $qq\b\q$ tetraquarks are determined by the action of the Hamiltonian on the $qq\q$ cluster \cite{Hogaasen:2004pm,Liu:2016ogz}. Each of the $qq\q$ configurations with spin $S$ yields a degenerate doublet of $qq\b\q$ states with spins $S+1/2$ and $S-1/2$. 

Note that the existence of degenerate doublets is totally independent of the nature of the Hamiltonian forming the $qq\bar q$ eigenstates. In particular, it is irrelevant whether the color triplet $qq\bar q$ eigenstate has $qq$ in color $\mathbf{\bar 3}$, color $\mathbf 6$, or a mixture of the two. Consequently degenerate doublets (in the heavy quark limit) occur regardless of whether the model uses the truncated~\cite{Wang:2016tsi} or full color basis (with~\cite{Stancu:2016sfd}  or without~\cite{Liu:2016ogz} mixing).

In the truncated color spectrum (for which color labels are superfluous) the states are classified according to the spin $s$ of the $qq$ pair, and the total spin $S$ of $qq\q$. The three possibilities for $(s,S)$ are $(0,1/2)$, $(1,1/2)$ and $(1,3/2)$. The hyperfine term in general mixes the $(0,1/2)$ and $(1,1/2)$ configurations, but this mixing disappears if the coefficients $\alpha_{ij}$ are independent of quark flavour \cite{Hogaasen:2004pm}. Either way, there are two  $qq\bar q$ eigenstates with $S=1/2$, each of which yields a degenerate $0^+/1^+$ doublet of $\subd$ states. Similarly the $(1,3/2)$ cluster forms a degenerate $1^+/2^+$ doublet. Deviating from the heavy quark limit breaks this degeneracy and states with the same total spin mix.

Models which include all color combinations, with or without mixing, have twice as many degenerate doublets (in the heavy quark limit). The four $qq\bar q$ eigenstates with $S=1/2$ yield four $0^+/1^+$ doublets, and the two $qq\bar q$ eigenstates with $S=3/2$ yield two $1^+/2^+$ doublets. For models without color mixing, the classification of states, and the degeneracy within levels, is discussed in ref.~\cite{Liu:2016ogz}. Incorporating color mixing changes the masses and color wavefunctions of the doublets, but leaves their
 degeneracy intact.

To illustrate the generic features of the spectra in such models, we consider the Hamiltonian of ref.~\cite{Wang:2016tsi} in the heavy quark limit (switching off all hyperfine interactions with $\bar b$) and with $SU(3)$ flavour symmetry in the hyperfine couplings. Using the notation of ref.~\cite{Wang:2016tsi}, the masses are controlled by the parameters $\kappa_{qq}$ and $\kappa_{q\bar q}$, which play the role of the $\alpha_{ij}$ coefficients in our notation but are normalised according to the color channels. From the Hamiltonian matrices in ref.~\cite{Wang:2016tsi} one readily obtains the mass formulae
\begin{align}
M_{0^+}=M_{1^+}&=M-\frac{3}{2}\kappa_{qq},\label{diq:1}\\
M_{0^+}'=M_{1^+}'&=M+\frac{1}{2}\kappa_{qq}-2\kappa_{q\bar q},\label{diq:2}\\
M_{1^+}''=M_{2^+}&=M+\frac{1}{2}\kappa_{qq}+\kappa_{q\bar q}.\label{diq:3}
\end{align}
Each of the $\kappa$ coefficients is a positive number, so for any choice of parameters the lightest pair is a $0^+/1^+$ doublet and the heaviest is a $1^+/2^+$ doublet. 

If the $\kappa$ coefficients are inferred from one gluon exchange, they are related $\kappa_{q\bar q}=\kappa_{qq}/2$. (This follows from their definition in ref.~\cite{Maiani:2004vq}, where $\kappa_{q\bar q}$ is normalised to a weighted combination of color singlet and octet interactions.) In this case the splitting among the states is controlled by a single parameter. The corresponding spectrum is depicted in the left panel of Fig.~\ref{fig-spectrum}.
}

The above considerations explain the spectrum in the diquark model of ref.~\cite{Wang:2016tsi}, where  three {approximately degenerate doublets} can be clearly seen. To characterise this as a diquark model does not seem appropriate, since the Hamiltonian is (approximately) diagonal in the basis where the spins of the light quarks $qq\q$, not the spins of diquarks  $qq$ and $\b\q$,  are good quantum numbers. Similar remarks apply to diquark models in other contexts, such as $Qq\Q \q$, where the spectrum is largely determined by the spin of $q\q$, rather than the spins of the diquarks $Qq$ and $\Q\q$. 

A different implementation of the diquark idea considers in equation (\ref{chromo}) only spin interactions within a diquark, ignoring those between quarks in different diquarks \cite{Maiani:2014aja,Lebed:2016yvr}. {(After submission of this paper, this approach was applied to $\X$ in ref.~\cite{Ali:2016gdg}.)} This is more consistent with the idea of diquarks as effective degrees of freedom, since it is equivalent to 
\begin{equation}
H=\sum_k m_k,
\end{equation}
where now $m_k$ are the masses of the constituent diquarks after spin splitting, with the axial somewhat heavier than the scalar. {The masses and corresponding $J^P$ quantum numbers of different combinations of diquarks $qq$ (scalar $S$ and axial $A$) and antidiquarks $\b\q$ ($\bar S$ and $\bar A$) are then
\begin{align}
{S\bar S:}&&M_{0^+}&=M,\label{mass:1}\\
{S\bar A:}&&M_{1^+}&=M+\delta,\\
{A\bar S:}&&M_{1^+}'&=M+\Delta,\\
{A\bar A:}&&M_{0^+}'=M_{1^+}''=M_{2^+}&=M+\delta+\Delta,\label{mass:4}
\end{align}
where $\Delta$ and $\delta$ are the mass differences between scalar and axial $qq$ and $\b\q$ states respectively. Note that  $\Delta>>\delta$, since the splittings scale inversely with quark masses, so there is a small separation between the lightest states $0^+$ and $1^+$, a larger separation to the next $1^+$ state, and higher still a degenerate triplet $0^+$, $1^+$, $2^+$. (In the model of ref.~\cite{Ali:2016gdg}, $\delta=50$~MeV and $\Delta=400$~MeV.)} The spectrum in this approach is summarised in the right panel of Fig.~\ref{fig-spectrum}. {The expectation in the diquark model of a $0^+/1^+$ pair with similar masses was noted in ref.~\cite{Lebed:2016yvr}.}


As shown in Fig.~\ref{fig-spectrum}, there are characteristic differences in the mass spectra of tetraquark models depending on whether diquarks are genuine effective degrees of freedom, or instead, there are pair-wise interactions among all quark constituents. If partners to $\X$ are eventually discovered, the pattern of their masses can be used to constrain models.

{ A feature common to both approaches is that the lightest states are an approximately degenerate $0^+/1^+$  pair, with exact degeneracy in the heavy quark limit. (As noted previously, 
models with pair-wise interactions among all quark constituents also produce degenerate $0^+/1^+$ doublets, regardless of whether the full or truncated color basis is used, and whether color mixing is or is not allowed.) Assuming that $\X$ belongs to this lightest $0^+/1^+$  doublet, it should therefore have a partner nearby in mass with different spin.} 

In particular, if $\X$ is itself an $I(J^P)=1(0^+)$ state (decaying to $B_s\pi$), it must have a $1(1^+)$ partner nearby (decaying to $B_s^*\pi$ with less phase space, hence narrow). Alternatively if $\X$ is a $1(1^+)$ state (decaying to $B_s^*\pi$ with a hidden photon) it would have a $1(0^+)$ partner (decaying to $B_s\pi$). Note also that in either interpretation a further pair of degenerate isoscalar partners $0(0^+)$ and $0(1^+)$ is expected, both very narrow as they are stable to strong decays. (Alternatively the neutral partners could mix with the isovectors as described in the previous section.)

In order to discuss the higher-lying states we need an estimate of spin splittings, which vary from model to model. We base our discussion on the splittings of ref.~\cite{Wang:2016tsi}, and comment on the other models below. In their approach the separation between the heaviest and lightest $\subd$ tetraquarks is 235~MeV. Re-scaling their masses to identify the lightest as $\X$, the heaviest partners would have masses around 5800~MeV. In this case very few strong decay channels are available to the tetraquark family: only $B_s\pi$, $B_s^*\pi$, and possibly, for the heaviest of the states, $B\*\K$ are kinematically accessible. Note that we expect a smaller overall splitting in models where spin interactions are restricted to within, rather than between, diquarks, namely for spectra of the type depicted in the right panel of Fig. \ref{fig-spectrum}. In this case it is likely that all states are below $B\K$ threshold.
 
Since the $B_s\*\pi$ channels are accessible only to isovectors, we can conclude that most, if not all, of the isoscalar tetraquarks are stable to strong decay. Only those above $B\*\K$ threshold can decay strongly. This seems unlikely for the $0(0^+)$ state, which lies at the top of the spectrum only in the genuine diquark model, where smaller splittings overall are expected. In order for the $0(1^+)$ to decay it would have to be above the heavier $B^*\K$ threshold, requiring a stronger splitting than ref.~\cite{Wang:2016tsi}. The $0(2^+)$ couples to $B\*\K$, but only in D-wave, and with non-conservation of quark spin; we thus expect that it, too, would be very narrow.

In general the isovector states have more possibilities. The heavier among the $1(0^+)$ and $1(1^+)$ states can decay to $B_s\pi$ and $B_s^*\pi$ respectively, but with more phase space than $\X$ and its degenerate partner, so presumably they will be broader. If sufficiently heavy they could also decay to $B\K$ and $B^*\K$. The $1(2^+)$ would decay to $B_s\*\pi$, and possibly $B\*\K$, but only in D-wave and with non-conservation of quark spin, as described above.

The spectra of refs~\cite{Liu:2016ogz,Stancu:2016sfd} contain twice as many states. The heavier among these can access several strong decay channels, but many remain below $B\K$ threshold, and so will be narrow for the reasons described above.

Tetraquark interpretations can occasionally be applied to a particular state without invoking a proliferation of partners. This only works if, due to the pattern of strong decay channels, the candidate itself is uniquely stable. In the case of $\X$ the scenario is very different. Among the many partners with various spin and isospin, it is one of comparatively few which can decay strongly. The rest would be extremely narrow, and their discovery in weak or radiative decays would support the tetraquark hypothesis.

\section{Conclusions}

We have examined a number of possible explanations of the $X(5568)$ signal and find that none of them give a satisfactory description of the observed state. In particular, the location and shape of the line shape make a threshold explanation unlikely. Cusp models rely on nearby hadronic  channels -- the only available channel is $B_s^*\pi$. This gets the mass correct and predicts the quantum numbers of the $X$ as $J^P = 1^-$. However, this explanation requires P-wave rescattering with a flavour-blind interaction. Both of these are not preferred in conventional phenomenology. If the cusp explanation finds support, then a series of similar ``states" are expected. For example, a neutral cusp should appear in $B_s\pi^0$ approximately 5 MeV below the $X$. Similar cusps would also be expected in $B_sK$ near $B_s^*K$ and in $B\pi$ near $B^*\pi$.

Possible molecular explanations are necessarily similar to cusp models since both rely on nearby hadronic thresholds. The only reasonable channel is $B\K$ but this either requires a binding of more than 200 MeV and a weak coupling to $B_s\pi$ or a coupled $B_s\pi - B\K$ system with  unusual GGC interactions. Neither of these scenarios are likely. 

The failure of plausible weak coupling scenarios encourages speculation based on strong coupling tetraquark models. Unfortunately, these appear to suffer similar problems; namely all natural estimates (by which we mean those with mass predictions tuned to well-known hadronic resonances) yield  masses that are 500 MeV or higher than the $X$ mass. If one ignores this basic problem, then tetraquark models predict a pair of nearby neutral $X$ states that should be visible in $B_s\pi$. In fact, many spin and flavour analogue states are expected. A novel feature of many of these is that they will be very long lived, and therefore should  be readily seen in appropriate channels. If such states are observed it will likely revolutionise the current understanding of strongly coupled QCD phenomenology.

Given the difficulty in constructing a viable resonance or weak coupling model of the $X$, it is prudent in enquire into the robustness of the experimental signal. An immediate concern is that the background peaks under the resonance, and this is enhanced by the ``cone cut" employed by \d0. Indeed, approximately one half of the signal events can easily be absorbed into a slight adjustment of the background shape, significantly reducing the significance of the observation. Thus it is important that this shape be accurately obtained in the sidebands.

A more elaborate possible confounding issue involves missing hadrons. The \d0 detector cannot detect pions at low transverse momentum. This raises the possibility that the $B_s\pi$ system is actually produced in an electroweak decay (of, say, the $B_c$) with an undetected hadron. Integrating events over the unknown degrees of freedom can yield a peak in the $B_s\pi$ system with a typical hadronic width. For example, $B_c\to B_s \rho \to B_s \pi [\pi]$ naturally gives rise to a kink in the $B_s\pi$ spectrum near 5570 MeV. If this peak were to be ameliorated at higher invariant mass (due to hadronic form factors) then it is possible to generate a signal similar to that of the $X$. Testing this scenario will require careful simulation accounting for detector efficiencies and hadronic form factors.

In summary, no viable explanation of the $X(5568)$ is apparent. While we are aware of the dangers of making a ``failure of imagination" argument, this suggests that extensive follow-up experimentation is in order, both to verify the original signal and to search for the  many possible adjunct states.

{ 
\section*{Note added}
Following the submission of this preprint to the arXiv and journal, LHCb reported on the search for $\X$ in their larger data sample, finding no evidence for the state and setting upper limits on its production~\cite{LHCb:2016ppf}. 

There have in addition between several theory papers reaching conclusions similar to ours. Guo \textit{et al.}~\cite{Guo:2016nhb} argue using chiral symmetry that $\X$ is too light to be a plausible tetraquark candidate, and provide quantitative arguments against the  cusp scenario. Using a relativised quark model L\"u and Dong~\cite{Lu:2016zhe} find $\subd$ tetraquark masses are too heavy for $\X$. Albaladejo~\textit{et al.}~\cite{Albaladejo:2016eps} consider the $B_s\pi-B\bar K$ coupled-channel system and conclude that a pole at the appropriate mass requires unnatural parameters. Chen and Ping~\cite{Chen:2016npt} study the four-quark $\subd$ system in the chiral quark model, finding that the diquark-diquark configurations are too heavy for $\X$, and that molecular configurations are not formed. In a lattice study of $B_s\pi$ scattering, Lang~\textit{et~al.}~\cite{Lang:2016jpk} find no evidence for $X(5568)$.
}

\acknowledgments

We are grateful to Don Lincoln, Jim Mueller, and Daria Zieminska for discussions on the \d0   results{, to Marek Karliner and Tim Gershon for pointing out the LHCb results, and to Richard Lebed and Fl.~Stancu for discussions on theoretical aspects.}

\end{document}